\documentclass[10pt,aps,prl,twocolumn,superscriptaddress]{revtex4}
\usepackage[linktocpage=true, colorlinks=true, urlcolor=blue, linkcolor=blue, citecolor=blue]{hyperref}
\usepackage{graphicx}
\usepackage{amsmath,amssymb}
\usepackage{color}
\usepackage{url}
\usepackage{hyperref}
\usepackage{cleveref}
\usepackage{xfrac}
\usepackage{upgreek}
\usepackage{siunitx}
\newcommand*\diff{\mathop{}\!\mathrm{d}}

\graphicspath{{./Images/}}

\begin{document}

\title{Retinal blood flow imaging with combined full-field swept-source optical coherence tomography and laser Doppler holography}

\author{L\'eo Puyo}
\affiliation{Institute of Biomedical Optics, University of L\"ubeck, Peter-Monnik-Weg 4, 23562 L\"ubeck, Germany}
\affiliation{gl.puyo@gmail.com}

\author{Hendrik Spahr}
\affiliation{Institute of Biomedical Optics, University of L\"ubeck, Peter-Monnik-Weg 4, 23562 L\"ubeck, Germany}

\author{Clara Pf\"affle}
\affiliation{Institute of Biomedical Optics, University of L\"ubeck, Peter-Monnik-Weg 4, 23562 L\"ubeck, Germany}

\author{Gereon H\"uttmann}
\affiliation{Institute of Biomedical Optics, University of L\"ubeck, Peter-Monnik-Weg 4, 23562 L\"ubeck, Germany}
\affiliation{Medical Laser Center L\"ubeck GmbH, Peter-Monnik-Weg 4, 23562 L\"ubeck, Germany}
\affiliation{Airway Research Center North (ARCN), Member of the German Center for Lung Research (DZL), Wöhrendamm 80, 22927 Großhansdorf, Germany}

\author{Dierck Hillmann}
\affiliation{Institute of Biomedical Optics, University of L\"ubeck, Peter-Monnik-Weg 4, 23562 L\"ubeck, Germany}
\affiliation{Thorlabs GmbH, Maria-Goeppert-Straße 9, 23562 L\"ubeck, Germany}
\affiliation{Department of Physics and Astronomy, Vrije Universiteit Amsterdam, De Boelelaan 1081, 1081 HV Amsterdam, Netherlands}
\affiliation{d.w.a.hillmann@vu.nl}

\date{\today}

\begin{abstract}
Full-field swept-source optical coherence tomography (FF-SS-OCT) and laser Doppler holography (LDH) are two holographic imaging techniques presenting unique capabilities for ophthalmology. We report on interlaced FF-SS-OCT and LDH imaging with a single instrument. Effectively, retinal blood flow and pulsation could be quasi-simultaneously monitored. This instrument holds potential for a wide scope of ophthalmic applications.
\end{abstract}

\maketitle
In the past few years, holographic imaging schemes have enabled the development of new functional contrasts in ophthalmology. By combining swept-source optical coherence tomography with digital holography, full-field swept-source OCT (FF-SS-OCT, or holographic OCT), allows phase stable retinal imaging at ultrafast volumetric rates~\cite{Povavzay2006, Bonin2010, Auksorius2019Crosstalk, Valente2021}. The phase stability within a single volume enables digital aberration compensation while the phase stability over consecutive volumes allows hemodynamic measurements and the imaging of sensory retinal functions~\cite{Hillmann2016, SpahrHillmann2015, Hillmann2016PNAS}.
Laser Doppler holography (LDH) combines laser Doppler imaging with digital holography. Compared to standard full-field laser Doppler or laser speckle contrast imaging~\cite{SerovLasser2005, Sugiyama2010, Cho2020portable}, the interferometric detection in LDH enables full-field retinal blood flow imaging with higher temporal resolution~\cite{Puyo2018}. The phase sensitivity in LDH additionally enables the detection of the local axial direction of blood flow, i.e., color Doppler imaging~\cite{Puyo2021Directional}.
Here, we report on adapting an FF-SS-OCT setup to allow OCT and LDH imaging to be performed on a single instrument, which we used to monitor retinal blood flow and better understand the hemodynamic signals measured by both modalities.


\begin{figure}[t!]
\centering
\includegraphics[width = 1\linewidth]{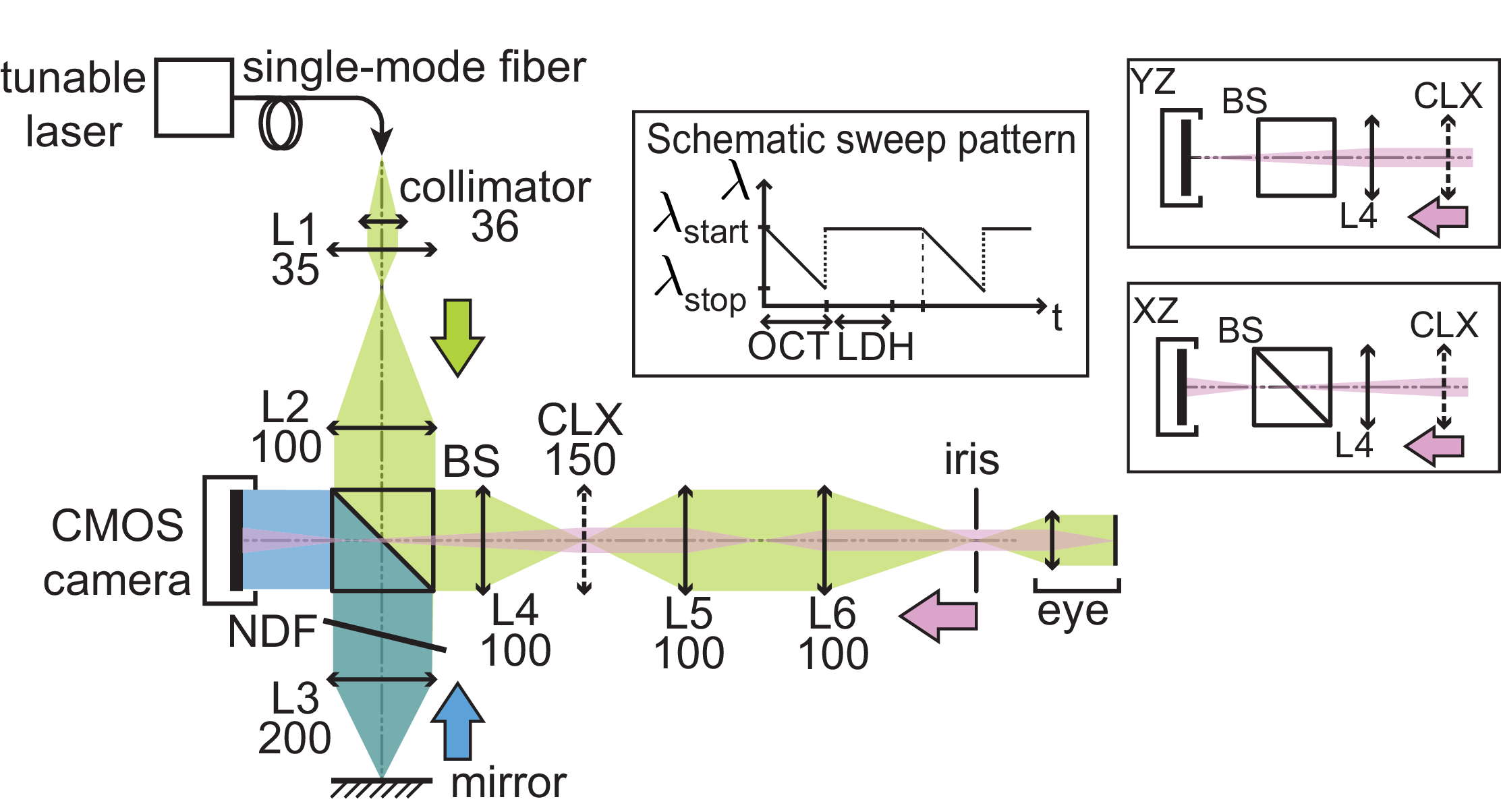}
\caption{Holographic OCT/LDH retinal imaging setup. BS: beam splitter, NDF: neutral density filter, CLX: cylindrical lens,
L1-L6: converging lenses with focal lengths given in millimeter below.
The light source alternates between a wavelength sweep for OCT and a stable wavelength for LDH. The cylindrical lens defocuses the retinal image on the camera along the horizontal direction.
}
\label{fig_1_Setup_HoloscopyDoppler}
\end{figure}

The interferometer used for retinal imaging is drawn in Fig.~\ref{fig_1_Setup_HoloscopyDoppler}. It differs from recently used FF-SS-OCT setups (\cite{SpahrHillmann2015, Hillmann2016, Hillmann2016PNAS})
mainly
by an additional cylindrical lens. The wavelength-tunable light source used for the experiments (Superlum BroadSweeper BS-840-1) offers a 51 nm tuning range with a central wavelength of 841 nm. The retina is illuminated with 5 mW over a disc of diameter 3.1 mm, and an area of size 2.6 mm $\times$ 1.5 mm was imaged onto the camera with a numerical aperture of 0.1. Investigations were done with 3 healthy volunteers; written informed consent was obtained from all subjects. Compliance with the maximum permissible exposure (MPE) of the retina and all relevant safety rules was confirmed by the responsible safety officer. All experiments were performed in accordance with relevant guidelines and regulations~\cite{ANSI_Z8036_2016}. The study was approved by the ethical board of the University of L\"ubeck (ethical approval "Ethik-Kommission L\"ubeck 16–080").
By default, after having performed its wavelength sweep, the laser goes back to the starting wavelength and remains stable until the start of the following cycle. We measured that the laser operates that transition in approximately 70 µs. An Arduino board triggers both the laser sweep and the camera batch recording. The stable wavelength part of the sweep cycle was used for LDH imaging. By changing the sweep-speed and sweep-range, the number of images recorded per cycle, and the time between two cycles, one can modify the number of images for the OCT and LDH series as well as the temporal resolution. The high-speed CMOS camera (Photron FASTCAM SA-Z) was typically used at a frame rate of
60 to 65 kHz
to record frames of size 640 $\times$ 368 pixels digitized with 8-bit.
A converging cylindrical lens acting on the X (horizontal) direction was added for the purpose of defocusing the holographic detection. As illustrated further, such a detection allows removing artifacts from LDH images and performing color Doppler imaging. This lens was placed in a plane conjugate to the tip of the single mode fiber so that the illumination beam was not significantly altered. The light backscattered by the eye was not affected in the Y (vertical) direction, and therefore remained focused onto the camera along the Y axis. However, along the X direction, the light converged so that the camera captured a retinal image horizontally defocused by approximately 5 cm. The cylindrical lens allowed preserving the live-preview of a B-scan based on the central lines of the ultrafast camera, similarly to other holographic ophthalmic setups~\cite{Ginner2018, Auksorius2021}.
The processing of OCT data was carried out with previously published algorithms, which includes aberration, dispersion, and axial motion compensation~\cite{Hillmann2019}. The digital aberration compensation and registration determined from the OCT data was also applied to LDH data. The main processing steps for LDH were singular value decomposition (SVD) filtering~\cite{Puyo2020}, and a Fourier analysis~\cite{Puyo2018}. Blood flow is computed as $M_1 = \int_{-\infty}^{+\infty} S(f) |f|   \diff f$, where $S$ is the SVD filtered power spectrum density and $f$ is the frequency.
Images are corrected for the non-uniform intensity distribution. The average of the images was then subtracted to remove the background signal.



\begin{figure}[t!]
\centering
\includegraphics[width = 1\linewidth]{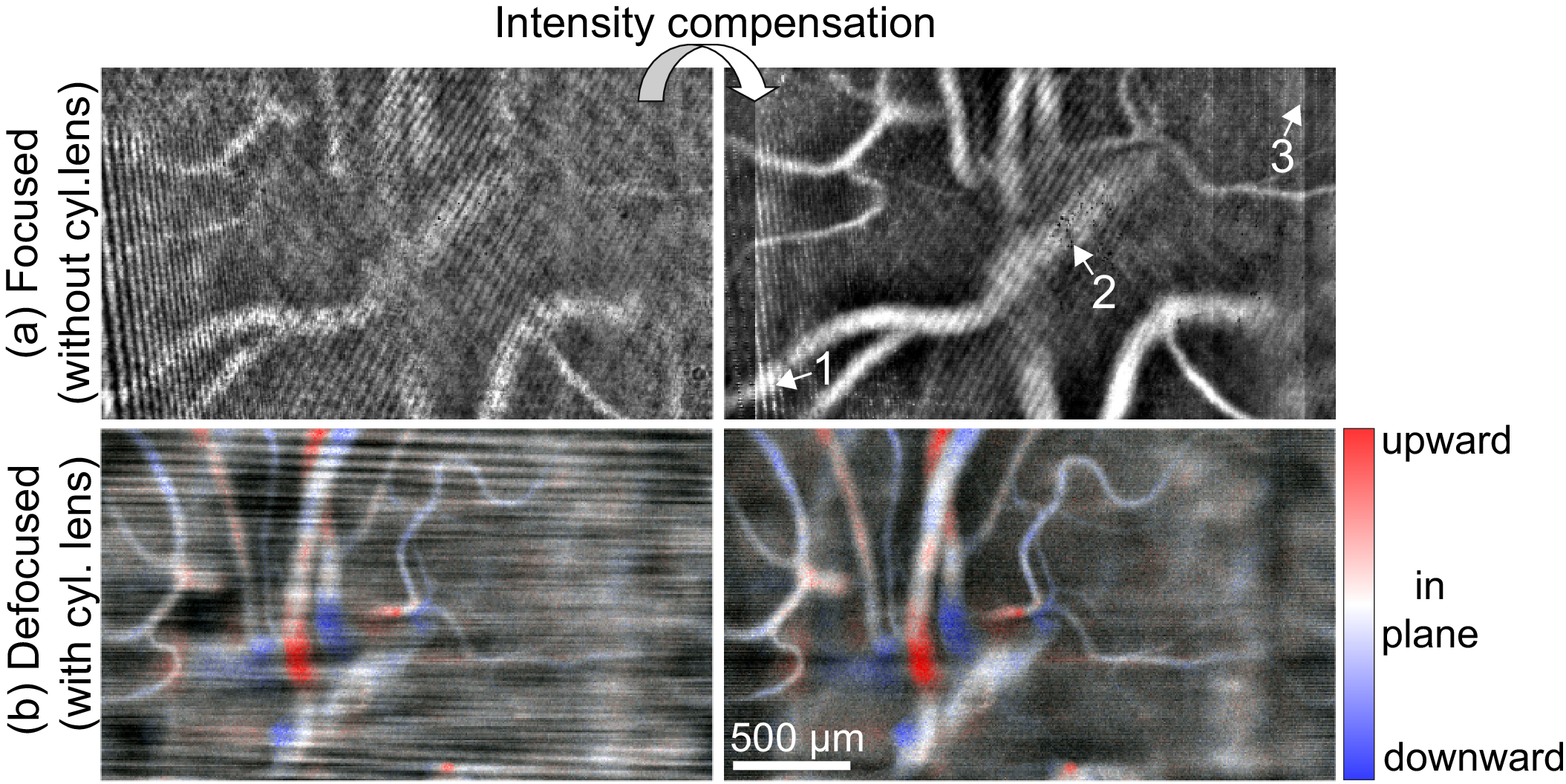}
\caption{
LDH, intensity correction and defocused measurement.
(a) Focused detection, the intensity correction improves the contrast but artifacts remain.  \textcolor{blue}{\href{https://opticapublishing.figshare.com/articles/media/Visualization_1_mp4/17207060}{Visualization 1}}.
(b) Defocused detection, the refocusing blurs the interferogram imprint and color Doppler imaging becomes possible. \textcolor{blue}{\href{https://opticapublishing.figshare.com/articles/media/Visualization_2_mp4/17207069}{Visualization 2}}.
}
\label{fig_2_WithWithoutASP}
\end{figure}

Figure~\ref{fig_2_WithWithoutASP} presents two LDH measurements performed close to the optic disc of the same eye without (upper row) and with (lower row) the cylindrical lens. The measurements were performed with a fixed wavelength, and in each configuration 2048 raw interferograms were used to compute the blood flow images. In the focused configuration in Fig.~\ref{fig_2_WithWithoutASP}(a), the initial image is severely influenced by the intensity distribution of the interferogram. The intensity $I$ measured on the camera results from the interference between the reference field $E_{\rm LO}$ (local oscillator, LO) and object beam $E$: $I = \lvert E  +  E_{\rm LO} \rvert^{2}$. The DC terms are filtered in the Doppler analysis, and blood flow images are derived from one of the two cross-terms $E_{\rm LO}  E^{*} + E_{\rm LO}^{*}  E$, where $^{*}$ stands for the complex conjugate. Therefore, the response of a given pixel to variations of object field $E$ is modulated by the local reference field $E_{\rm LO}$.
We numerically corrected LDH interferograms to compensate for the pixel-to-pixel differences in the blood flow response caused by the non-uniform camera illumination. The intensity on the camera is considered approximately equal to the reference beam intensity, i.e., $I\approx \lvert E_{\rm LO} \rvert^{2}$. For each short-time window, we computed pixel-wise the corrected intensity $I_{\rm corr} = I / \sqrt{\overline{I}}$, where $\overline{I}$ is the temporal mean of $I$. This operation allows the temporal Fourier analysis of $I_{\rm corr}$ to reveal the Doppler induced fluctuations of the local object field $E$ independently from the pixel illumination by the reference field $E_{\rm LO}$.
As shown in Fig.~\ref{fig_2_WithWithoutASP}, this intensity correction dramatically improves the quality of blood flow images, but there remain artifacts in the focused configuration, particularly visible with eye movements in \textcolor{blue}{\href{https://opticapublishing.figshare.com/articles/media/Visualization_1_mp4/17207060}{Visualization 1}}. Fringes from the camera illumination pattern (interferogram imprint) are still visible. Overly bright pixels are seen (arrow '1'); these were insufficiently illuminated and thus more affected by camera readout noise. Other pixels are black because they were saturated and could not detect Doppler fluctuations (arrow '2').
Finally, other structures resulting from spatially varying pixel responses are also visible (arrow '3').
Overall, these artifacts correspond to areas where the approximation $I\approx \lvert E_{\rm LO} \rvert^{2}$ is invalid, leading to residual pixel-to-pixel differences in blood flow response.
Figure~\ref{fig_2_WithWithoutASP}(b) shows a measurement in the same region after adding the cylindrical lens in the setup. In presence of the lens, a numerical propagation is made along the X direction to propagate the hologram into focus. This propagation blurs the camera-plane pixel-to-pixel differences in blood flow sensitivity because they become out-of-focus. Residual artifacts due to the interferogram illumination and the camera readout noise are effectively blurred out. The situation is comparable to having dust on lenses that is defocused and therefore becomes unnoticeable in the image plane. The removal of the interferogram imprint from Doppler images can be well appreciated in \textcolor{blue}{\href{https://opticapublishing.figshare.com/articles/media/Visualization_2_mp4/17207069}{Visualization 2}}.
The second advantage of the numerical propagation is that reconstructed holograms are complex-valued so their temporal Fourier transform can be asymmetrical. Positive and negative Doppler frequency shifts can only then be distinguished
enabling the directional information in Fig~\ref{fig_2_WithWithoutASP}(b)~\cite{Puyo2021Directional}.
With a hue/saturation/value combination of images based on the symmetry/asymmetry of the power spectrum density, directional blood flow images can be obtained~\cite{Puyo2021Directional}. Therefore, the defocused measurement allows vessels carrying upward or downward flow to be revealed with a red and blue contrast.

\begin{figure}[t!]
\centering
\includegraphics[width = 1\linewidth]{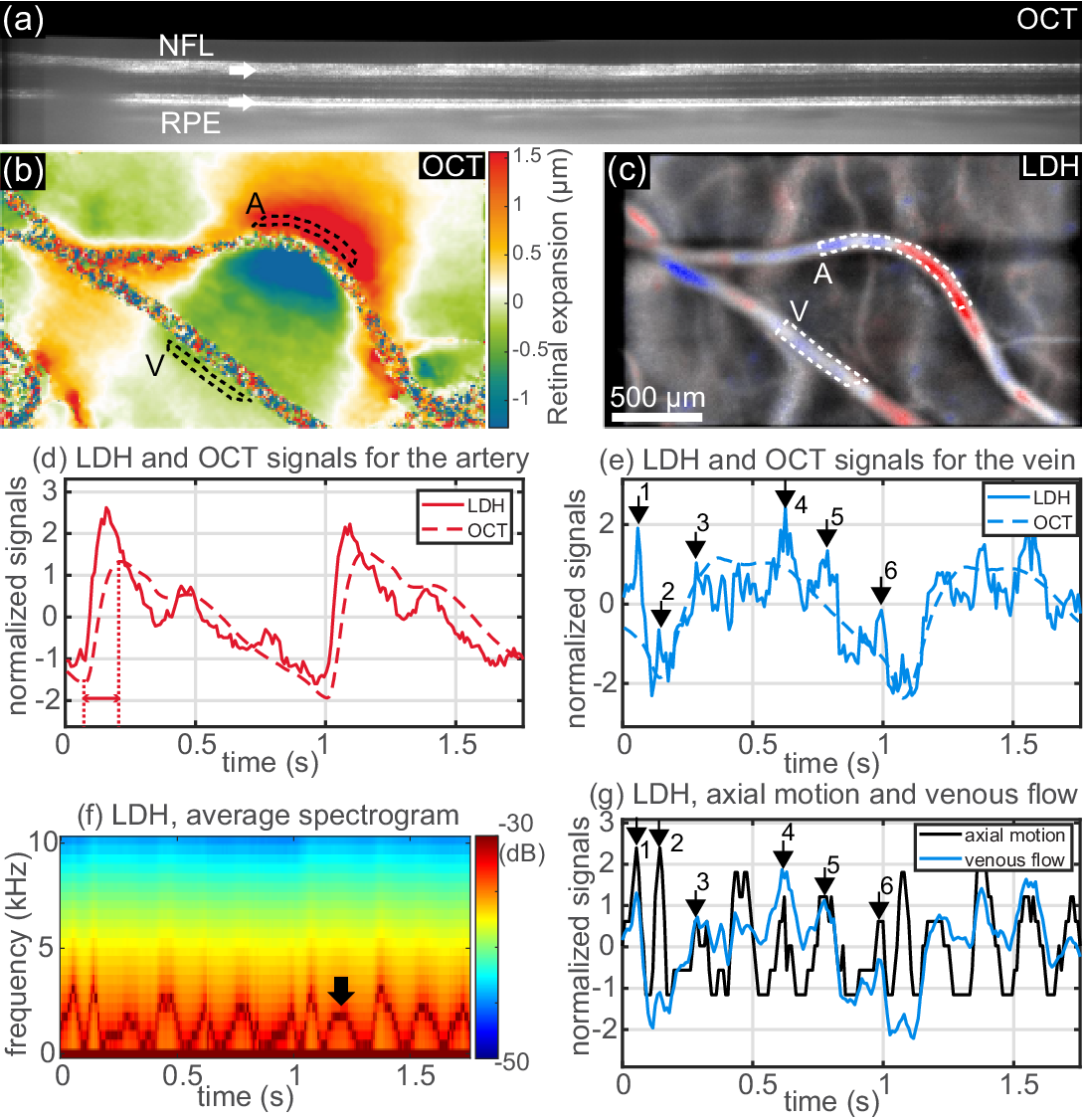}
\caption{LDH/OCT retinal blood flow imaging.
(a) OCT B-scan.
(b) Retinal thickness changes.
(c) Directional blood flow.
(d-e) Blood flow and retinal thickness for the artery and vein.
(f) LDH, spectrogram showing the axial eye motion.
(g) LDH, correlation between axial motion and venous flow artifacts.
\textcolor{blue}{\href{https://opticapublishing.figshare.com/articles/media/Visualization_3_mp4/17207072}{Visualization 3}}.
}
\label{fig_4_OCT_Doppler}
\end{figure}

We used quasi-simultaneous OCT/LDH measurements to visualize retinal blood flow dynamics in Fig.~\ref{fig_4_OCT_Doppler}.
OCT volume-to-volume variations of phase difference between the retinal pigment epithelium (RPE) and nerve fiber layer (NFL) were smoothed, unwrapped, and integrated to reveal local retinal thickness changes. It was previously shown that the blood pulsation induces thickness changes that can be monitored with an accuracy that was estimated to 10 nm~\cite{SpahrHillmann2015}. This method was used to measure the pulse wave velocity in retinal arteries~\cite{SpahrHillmann2015}, and recently implemented with scanning-instruments~\cite{Desissaire2021}.
The map of the local changes of retinal thickness between the end of diastole and the following systolic peak (time points indicated by the double headed arrow in Fig.~\ref{fig_4_OCT_Doppler}(d)) is shown in Fig.~\ref{fig_4_OCT_Doppler}(b). The pulse wave has not reached the venous vasculature yet, therefore the image reveals the retinal expansion induced by arteries only. The phase is scrambled in vessels due to the rapid phase decorrelation induced by blood flow. For this type of measurement, the axial resolution only needs to be high enough to separate the NFL and RPE layers. The laser was swept only along 20 nm instead of the full available range to obtain A-scans with a larger depth range at a reduced axial resolution.
Two hundred cycles consisting of 200 images for OCT and 150 images for LDH were recorded at 65 kHz. It is critical that the temporal variations of the RPE-NFL phase difference are sampled properly to avoid phase wrapping; the time between two volumes was here 9 ms. Volumes affected by micro-saccades were manually removed (typically 3 consecutive volumes were rejected for each micro-saccade).
The directional blood flow image shown in Fig.~\ref{fig_4_OCT_Doppler}(c) was obtained from the LDH data.
In Fig.~\ref{fig_4_OCT_Doppler}(d-e), the changes of retinal thickness measured by OCT in the vicinity of an artery (red, 'A') and vein (blue, 'V') are compared to the blood flow measured by LDH in the same vessels. Plots were normalized to have a 0 mean and a standard deviation of 1. The non-normalized graphs and corresponding movies are shown in \textcolor{blue}{\href{https://opticapublishing.figshare.com/articles/media/Visualization_3_mp4/17207072}{Visualization 3}}.
The retinal thickness changes near the artery and vein closely match LDH blood flow measurements in those same vessels; we observed this in all 3 healthy subjects we imaged. On closer look, it can however be seen that the LDH arterial signal is slightly earlier than its OCT counterpart. More importantly, the correlation between OCT and LDH is not perfect, particularly for the venous signal (arrows in Fig.~\ref{fig_4_OCT_Doppler}(e)). The LDH average spectrogram in Fig.~\ref{fig_4_OCT_Doppler}(f) allows monitoring the eye axial velocity (arrow)~\cite{Puyo2021Directional}. Bulk motion is significantly filtered in LDH thanks to the SVD filtering~\cite{Puyo2020}. However, when comparing the smoothed venous blood flow measured by LDH to the axial bulk motion in Fig.~\ref{fig_4_OCT_Doppler}(g), it appears that the time points where the LDH venous blood flow differs from the retinal thickness changes near the vein coincide exactly with bulk motion extrema. Therefore, these seemingly pulsatile artifacts are more likely caused by heartbeat-induced axial motion of the whole head than by retinal blood flow~\cite{Kinkelder2011}. These artifacts are much less visible in the arterial trace because of the larger amplitude of arterial flow variations.

\begin{figure}[t!]
\centering
\includegraphics[width = 1\linewidth]{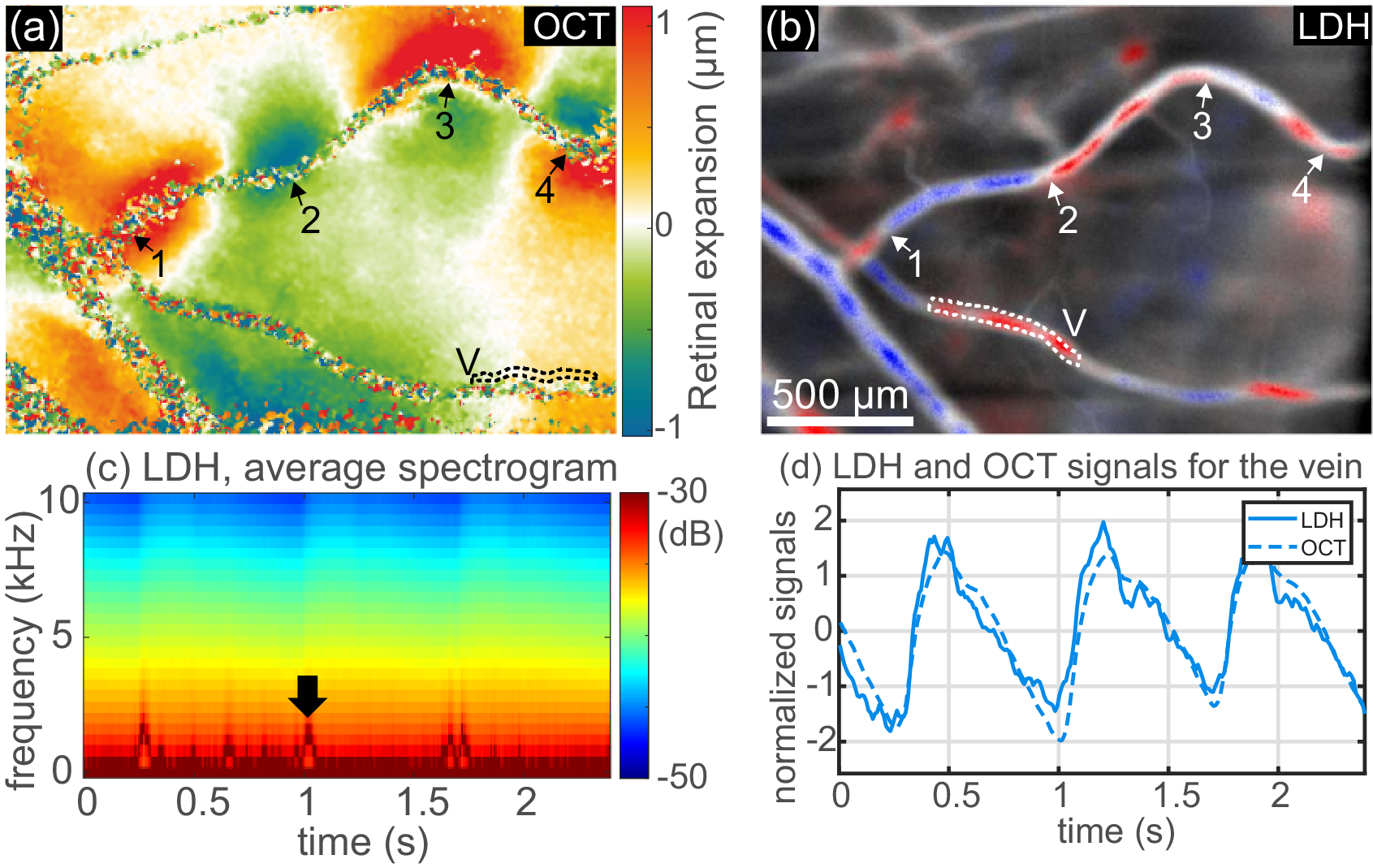}
\caption{Tissue expansion and vessels' in and out-of-plane geometry.
(a) Retinal expansion.
(b) Directional blood flow.
(c) LDH average spectrogram.
(d) LDH venous blood flow and OCT-measured retinal thickness changes.
\textcolor{blue}{\href{https://opticapublishing.figshare.com/articles/media/Visualization_4_mp4/17207078}{Visualization 4}}.
}
\label{fig_5_OCT_Doppler_directional}
\end{figure}

Figure~\ref{fig_5_OCT_Doppler_directional} shows another dual OCT/LDH measurement which illustrates how the RPE-NFL thickness variations are influenced by the local curving of nearby blood vessels. This measurement was realized with a face mask instead of a chinrest; 240 volumes consisting of 150 images for OCT and another 150 images for LDH were recorded at 60 kHz (10 ms between two cycles). The retinal expansion between the end of diastole and the following systolic peak is shown in Fig.~\ref{fig_5_OCT_Doppler_directional}(a). The directional blood flow image in Fig.~\ref{fig_5_OCT_Doppler_directional}(b) allows identifying local axial curving extrema along the blood vessels. An axial elevation of a vessel is for example evidenced by a section with a red contrast followed by a section with a blue contrast.
It is known, that when vessels are curved, the force exerted by the flow pressure is asymmetrically applied to the vessels walls~\cite{Han2012}. A greater force is exerted on the wall exposed to the incoming flow; therefore it is more subject to dilation or slight displacement.
The arrows '1' and '2' along the artery show that in the vicinity of a vessel section where the axial curving has reached an upward or downward local curving extrema, the retinal thickness on both sides of the vessel is increased or decreased, respectively. We hypothesize that when flow reaches an upwards maximum (arrow '1'), the pressure is exerted on the superior vessel wall, the NFL is pushed upwards, the distance between NFL and RPE is increased, and the corresponding phase shift is positive. Conversely, in the vicinity of a vessel section with downward flow (arrow '2'), the NFL-RPE phase shift is negative, presumably because the vessel is entangled in retinal tissue and therefore presses the NFL closer to the RPE.
The situations in the regions pointed by the arrow '3' and '4' are slightly different, as in these cases there is an in-plane curving that seems to have a greater influence than the out-of-plane one. Thickness changes of opposite signs are observed on the left and right sides of the artery. We assume that on the vessel's exterior side, the NFL is pushed upwards, resulting in an increase of the NFL-RPE thickness. Conversely, on the vessel's interior side the suction force caused by the vessel's slight displacement attracts the NFL downwards, resulting in a thickness decrease. It is also expected that the volume conservation and incompressibility of the liquid containing tissue between the RPE and NFL comes into play at the scale of the whole fundus. In practice, both the in-plane and out-of-plane curvings of vessels influence the local retinal expansion, adding up their effects. For example, in the area pointed by the arrow '2', a decrease of the NFL-RPE on both side of the vessel is expected from the axial curving revealed by the color Doppler image, but the slight in-plane curving induces an increase of retinal thickness on the exterior side of the vessel. This results in a decrease of thickness on the interior side of the vessel and almost no change of thickness on the exterior side.
Finally, in Figs.~\ref{fig_5_OCT_Doppler_directional}(c-d) shows the LDH average spectrogram and the comparison between the venous blood flow measured by LDH and the retinal thickness changes near the vein. The regions of interest were selected to avoid visible underlying choroidal vessels for LDH, and nearby blood vessels of opposite type for OCT. These graphs illustrate that when axial motion of the head is reduced thanks to the face mask (arrow in Fig.~\ref{fig_5_OCT_Doppler_directional}(c)), a better correlation is found between blood flow measured by LDH and retinal thickness changes.

We have shown that holographic OCT and LDH can be combined on one single setup. Axial resolution was reduced to perform blood flow imaging with OCT, but full resolution morphological images can be attained by changing the sweep pattern.
The numerical propagation along one direction in LDH is sufficient to obtain the directional blood flow contrast, and blurs the pixel-to-pixel variations in blood flow response. In contrast with previous LDH demonstrations~\cite{Puyo2018}, we did not use cross-polarized light; corneal specular reflections were not found to be an issue thanks to the limited coherence length of the laser (approximately 7 mm). The resulting instrument preserves the key features of both imaging techniques, yielding an instrument capable of imaging both blood flow and the 3D structure of the eye fundus. The OCT and LDH data can be straightforwardly corrected for the same aberrations and precisely co-registered.
One limitation for the instrument is that a factor of two is lost for the attainable temporal resolution (for a same number of images per cycle for OCT and LDH). The resulting temporal resolution remains however sufficient for most applications of OCT and LDH. For OCT, one limitation is that we do not measure the pressure directly, but instead the tissue displacement induced by it. Also, we observed that there can be a cumulative effect of retinal expansions in regions with nearby vessels of different types (artery/vein). Finally, the required fast volumetric rate places a significant constraint, and here, we were limited by the camera's on-board memory of 16 GB. For our frame rate and frame size, this memory allows a continuous measurement duration of about 1.1 s, which is not fully satisfying to image blood flow dynamics since it is approximately the duration of one cardiac cycle. To allow the measurement of several consecutive cardiac cycles, the duty cycle as well as the number images recorded for OCT and LDH series were reduced to numbers just sufficient for the functional measurements we investigated.
One implication of the fact that retinal tissue around blood vessels matches the waveform of blood flow passing through those vessels is that OCT should be usable to measure the pulse wave propagation velocity~\cite{SpahrHillmann2015}. These combined imaging modalities improved our understanding of the expansion of retinal tissue induced by blood vessels' pulsation. We observed that the in-plane and out-of-plane curving of vessels amplifies the local retinal tissue displacement and determines whether there the local NFL-RPE thickness becomes thinner or thicker. OCT and LDH provide complementary information about retinal blood flow and pulsation. The latter can be used to calculate the resistivity index as it requires information about the offset flow value, which is known in LDH but not with OCT. For its part, OCT provides aberration correction, is able to evidence axial motion-induced artifacts in LDH blood flow traces, and brings a quantitative information about the tissue expansion. Combining these information of blood flow and vessels deformation could bring valuable information about vessels elasticity. Finally, although we exemplified the value of combining OCT with LDH by investigating the biomechanical effects of retinal blood flow, this combination is promising for other ophthalmic applications such as the study of neuro-vascular coupling~\cite{Hillmann2016PNAS, Garhofer2020retinal}.

In conclusion, we have demonstrated hybrid OCT/LDH holographic retinal imaging, which we used to show that the thickness of the retinal tissue surrounding blood vessels reflects the blood flow passing through these vessels.
Overall, this instrument holds potential for a wide scope of applications in ophthalmology, ranging from basic research to clinical diagnosis.

\section*{Funding}
Deutsche Forschungsgemeinschaft (HU 629/6-1).

\section*{Disclosures}
DH: Thorlabs GmbH (E).

\section*{Data availability}
Data underlying the results presented in this paper are not publicly available at this time but may be obtained from the authors upon reasonable request.


\section*{Supplementary Material}
\vspace{-1em}
\noindent

\begin{center}
\textcolor{blue}{\href{https://youtu.be/TDqk9vg9a04}{Video Presentation}} \\
\end{center}

\bibliographystyle{unsrt}
\bibliography{./Bibliography}

\end{document}